\documentclass{sf2a-conf2018}
\usepackage[utf8]{inputenc}
\usepackage{graphicx}
\usepackage{hyperref}
\usepackage[]{natbib}  
\usepackage[dvipsnames]{xcolor}
\usepackage{amsmath}
\usepackage{wasysym}
\usepackage{pifont} 
\usepackage[referable]{threeparttablex}
\usepackage{booktabs, ltablex}
\keepXColumns
\usepackage{tabularx}

\def\BibTeX{{\rm B\kern-.05em{\sc i\kern-.025em b}\kern-.08em
    T\kern-.1667em\lower.7ex\hbox{E}\kern-.125emX}}
\bibpunct{(}{)}{;}{a}{}{,}  
\renewcommand{\b}[1]{{\bf#1}}
\newcommand{\bs}[1]{{\boldsymbol#1}}
\newcommand{\h}{\mathrm{nw}}
\newcommand{\Le}{\mathrm{Le}}
\newcommand{\Ro}{\mathrm{Ro}}
\newcommand{\Roh}{\mathrm{Ro}_\mathrm{t}}

\newcommand{\KE}{\mathrm{KE}}
\newcommand{\ME}{\mathrm{ME}}
\newcommand{\lc}{l_\mathrm{c}}
\newcommand{\uc}{u_\mathrm{c}}

\definecolor{darkorange}{HTML}{FF8C00}
\definecolor{jv}{HTML}{BFBF00}

\begin{document}

\TitreGlobal{SF2A 2018}


\title{Does magnetic field modify tidal dynamics\\ in the convective envelope of Solar mass stars?}

\runningtitle{Astoul et al.}

   \author{A. Astoul}\address{AIM, CEA, CNRS, Université Paris-Saclay, Université Paris-Diderot, Sorbonne Paris Cité, F-91191 Gif-sur-Yvette, France}
   \author{S. Mathis$^1$}
   \author{C. Baruteau}\address{IRAP, Observatoire Midi-Pyrénées, Université de Toulouse, 14 avenue Edouard Belin, 31400 Toulouse, France}
   \author{F. Gallet}\address{Université Grenoble Alpes, CNRS, IPAG, 38000 Grenoble, France}
    \author{K.\,C. Augustson$^1$}
   \author{E. Bolmont$^1$}\address{Department of Astronomy, University of Geneva, Chemin des Maillettes 51, 1290 Versoix, Switzerland}
    \author{A.\,S. Brun$^1$}
    \author{A. Strugarek$^1$}

\setcounter{page}{237}
\maketitle

\begin{abstract}
The energy dissipation of wave-like tidal flows in the convective envelope of low-mass stars is one of the key physical mechanisms that shape the orbital and rotational dynamics of short-period planetary systems. Tidal flows, and the excitation, propagation, and dissipation of tidally-induced inertial waves can be modified by stellar magnetic fields (e.g., Wei 2016, 2018, Lin and Ogilvie 2018). It is thus important to assess for which stars, at which location of their internal structure, and at which phase of their evolution, one needs to take into account the effects of magnetic fields on tidal waves. Using scaling laws that provide the amplitude of dynamo-generated magnetic fields along the rotational evolution of these stars (e.g., Christensen et al. 2009, Brun et al. 2015), combined with detailed grids of stellar rotation models (e.g., Amard et al. 2016), we examine the influence of a magnetic field on tidal forcing and dissipation near the tachocline of solar-like stars. We show that full consideration of magnetic fields is required to compute tidal dissipation, but not necessarily for tidal forcing.
\end{abstract}

\begin{keywords}
Magnetohydrodynamics -- waves -- planet-star interactions -- stars: evolution -- stars: rotation -- stars: magnetic fields
\end{keywords}

\section{Introduction}
Star-planet tidal interactions are thought to play a key role in the dynamical evolution of close-in exoplanets, by circularizing their orbit, synchronizing the rotational and orbital periods, and possibly leading to some evolution of the spin-orbit (mis-)alignment angle \citep[e.g.,][]{M2013,A2012,BM2016,DM2018}. Tides generally fall into two categories \citep{Z1977,O2014}: a quasi-static equilibrium tide, which corresponds to the emergence of an equatorial bulge via large-scale tidal flows, and a non-static wave-like component called the dynamical tide, which is associated with energy dissipation of tidally-induced waves. 
Dynamical tides in low-mass stars manifest as inertial waves in their convection zone and as gravito-inertial waves in their radiative zone. The dissipation of these waves, via turbulent friction and thermal diffusion, is part of what shapes the rotational and orbital properties of a two-body system \citep{H1981}. 

A body of recent works have explored the impact of various physical processes on tidal dissipation in the convective envelope of late-type stars, such as differential rotation \citep{OL2004, BR2013, G2016a}, turbulent friction \citep{OL2012,M2016}, and the effects of stellar evolution and metallicity \citep{M2015,BM2016,G2017,BG2017}. These works have shown that tidal dissipation varies strongly with the star's and planet's physical properties. It is only recently that the question of how stellar magnetism influences the propagation and dissipation of tidal inertial waves has been addressed \citep{W2016,W2018,LO2018}. Low-mass stars, which have a radiative zone under the convective envelope (from $0.4$ to $1.4$ solar masses), host indeed a powerful dynamo that is sustained by turbulent convection and differential rotation in their envelope \citep{BB2017}. In this context, \citet{W2016} and \citet{LO2018} have shown that strong magnetic fields can significantly affect the propagation of inertial waves, and cause Ohmic dissipation of the (magneto-)inertial waves to dominate over their turbulent viscous dissipation. Still, the question of how magnetic fields impact the tidal forcing of inertial waves via the large-scale equilibrium tide flows remains an open question, which we address in this communication, and which is the subject of a paper in preparation by Astoul et al.

To perform this analysis, we compare the magnitudes of the Lorentz and Coriolis forces in the effective tidal forcing driven by the equilibrium tide flows, which features a dimensionless number called the Lehnert number \citep[it is the ratio of the Alfvén speed and the rotational velocity,][]{L1954}. Various scaling laws are explored to estimate the magnitude of the Lehnert number throughout the convective envelope, which depends on the surface rotation period of the star. A parametric study using the 1D stellar evolution code STAREVOL \citep{A2016} is carried out to evaluate the Lehnert number at the base of the convective envelope of a solar-mass star along its lifetime, which we apply to a few observed star-planet systems. The same parametric study is used to assess the relative magnitudes of the Ohmic dissipation and the viscous turbulent dissipation for inertial waves, which is also traced by the Lehnert number \citep{LO2018}.

\section{Impact of the star's magnetic field on the tidal forcing of inertial waves}
\label{sec:sect2}

\subsection{An analytical criterion}
In the linearised Navier-Stokes equation for the dynamical tide, the tidal force arises from an effective forcing driven by velocity and displacement fields associated with the equilibrium tide \citep[see Appendix B from][]{O2005,O2013}.
In the presence of a magnetic field, the Lorentz force comes into play, affecting both dynamical and equilibrium tides. 
\citet{LO2018} have investigated the action of uniform and dipolar magnetic fields on the propagation and dissipation of inertial waves, without taking into account, in their simulations, the impact of the magnetic field on the tidal forcing.
When doing so, the effective tidal force density is the sum of a classical hydrodynamical part $f_1$, due to the hydrostatic equilibrium tide in the fluid's rotating frame, and of the linearised Lorentz force density $f_2$ \citep[see Appendix B from][ Astoul et al. in prep.]{LO2018}:
\begin{equation}
\left\{\begin{aligned}
f_1=&-\rho(\partial_t\b u_\h+2\Omega\b e_z\wedge\b u_\h)\\
f_2=&\frac{\bs\nabla\wedge\b B}{\mu_0}\wedge\b b_\h+\frac{\bs\nabla\wedge\b b_\h}{\mu_0}\wedge\b B
\end{aligned}\right.,
\end{equation}
where $\b u_\h$ and $\b b_\h$ are the non wave-like flow and magnetic field (referring to the equilibrium tide), respectively, and $\rho$, $B$, and $\Omega$ denote the mean density, the magnetic field maximum amplitude ($\b B$ being the field itself), and the equatorial rotational frequency, respectively.
The non wave-like magnetic field $\b b_\h=\bs\nabla\wedge(\bs\xi_\h\wedge\b B)$, with $\bs\xi_\h$
 the equilibrium tide/non-wave like displacement, is deduced from the equation of induction, where we neglect the action of magnetic diffusivity.
The ratio of both components of the tidal forcing reads (Astoul et al. in prep.):
\begin{equation}
\frac{f_2}{f_1}=\mathcal{O}\left(\frac{\Le^2}{\Roh}\right)\ \text{ with }\ 
\left\{\begin{aligned} 
&\Le=\frac{B}{\sqrt{\rho\mu_0}2\Omega R}\\
&\Roh=\frac{|\sigma_\mathrm{t}|}{2\Omega}
\end{aligned}\right. ,
\label{equation:eq1}
\end{equation}
where we have introduced the Lehnert number $\Le$, the Doppler-shifted Rossby number $\Roh$, and the tidal frequency in the rotating frame $\sigma_\mathrm{t}$ (for its definition, see section \ref{sec:sys} in the case of circularised and synchronised systems). 
Estimate of the time-dependent ratio $\Le^2/\Roh$ will allow us to know how the forcing through the linearised Lorentz force induced by the equilibrium tide compares to the forcing through the Coriolis acceleration applied on this flow, and whether the Lorentz force needs to be taken into account in the tidal forcing of inertial waves.

\subsection{Estimate of the magnetic field with simple scaling laws}
\begin{figure}[ht!]
 \centering
 \includegraphics[width=0.49\textwidth,clip]{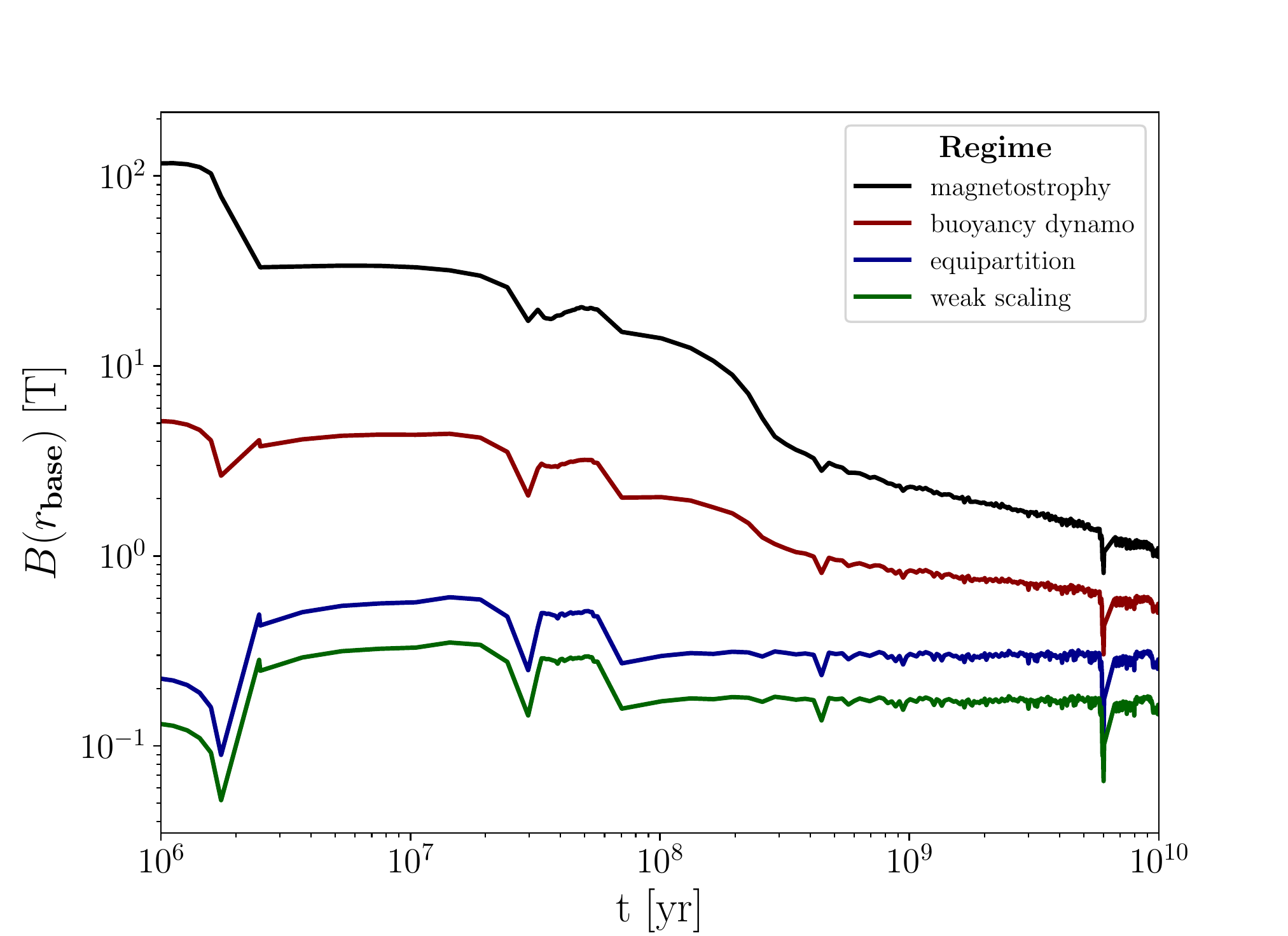}
 \includegraphics[width=0.49\textwidth,clip]{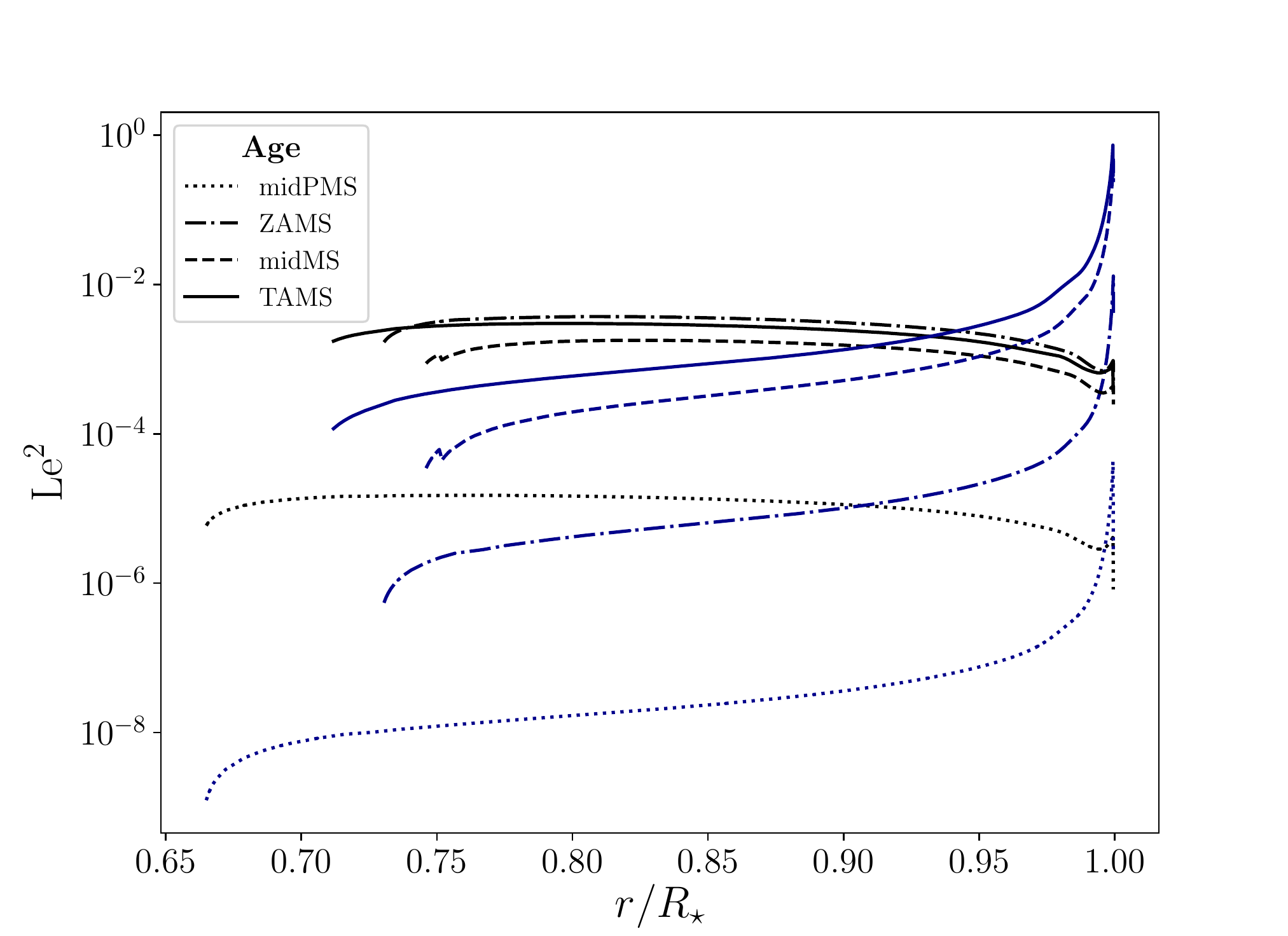}      
  \caption{{\bf Left:} Magnetic field (in Tesla) at the base of the convective zone against time for a $1M_\odot$ star, obtained from the different scaling laws introduced in Table~\ref{table:tab1}. 
  Results are obtained for a fast initial rotation ($P=1.6\mathrm{d}$).
{\bf Right:} Lehnert number squared against radius (normalized to the
star radius $R_{\star}$) in the magnetostrophic (\textit{black}) and equipartition (\textit{blue}) regimes, for different evolutionary stages: amid the pre-main sequence ("midPMS"; $\sim30\,\mathrm{Myr}$), at zero-age main sequence ("ZAMS"; $\sim70\,\mathrm{Myr}$), amid main sequence ("midMS"; $\sim5.3\,\mathrm{Gyr}$) and towards the end of the main sequence ("TAMS"; $\sim10.5\,\mathrm{Gyr}$).}
  \label{astoul:fig1}
\end{figure}

To estimate how the radial profile of the Lehnert number varies with time in the convective envelope of a $1M_\odot$ star, we have used the 1D stellar evolution code called STAREVOL \citep{A2016}. 
The code does not include the evolution of the star's magnetic field, which, however, may play a key role in transporting angular momentum, especially around the tachocline in solar-type stars \citep[e.g.][]{SB2011,BS2017}.
In table \ref{table:tab1}, we list simple prescriptions giving rough approximations of the magnetic field's strength as well as the Lehnert number expressed with quantities computed by STAREVOL (Astoul et al. in prep.). These dynamo prescriptions are based upon different reservoirs of magnetic energy: viscous diffusion for the weak scaling, kinetic energy for the equipartition, and gravitational energy for the buoyancy dynamo, or forces balance as is the case of the magnetostrophic regime where the Coriolis acceleration equates to the Lorentz force. In the Sun, the toroidal magnetic field at the tachocline is expected to be strong (about one Tesla, e.g. \citealp{C2013}), which is checked by the super-equipartition or the magnetostrophic regimes. This can be seen in the left panel of Fig.~\ref{astoul:fig1}, where we display the magnetic field at the base of the convective zone against time, deduced from the different scaling laws listed in Table~\ref{table:tab1} and applied thanks to our STAREVOL calculations. The buoyancy dynamo model works well for fast and young stars like T-Tauri stars, as well as fast rotating giant planets like Jupiter \citep{C2010}.

\begin{table}[h]
\centering
\begin{threeparttable}
\begin{TableNotes}[para,raggedright]
\item[1] \label{2} \cite{BG2015}  \item[2]\label{3}\cite{BB2017} \\
\item[3]\label{4}\cite{CN1986}  \item[4]\label{1}\cite{AMB2017} 
\end{TableNotes}
\begin{tabularx}{\textwidth}{c||cccc}
Regime & Weak scaling\tnotex{2} & Equipartition\tnotex{2}\tnote{,} \tnotex{3}  & Buoyancy dynamo\tnotex{1} & Magnetostrophy\tnotex{2}\tnote{,} \tnotex{4}\\
\hline
\rule[-2ex]{0pt}{6ex}  Balance & $ F_\mathrm{L}= F_\nu$ & $\ME=\KE$  & $\cfrac{\mathrm{ME}}{\mathrm{KE}}=\Ro^{-1/2}$ &  $F_\mathrm{L}=\rho a_\mathrm{c}$ \\
$\Le\times R/\lc$  & $\Ro/\sqrt{3}$  &  $\Ro$ & $\Ro^{3/4}$ & $\sqrt{\Ro}$\\
\insertTableNotes
\end{tabularx}
\end{threeparttable}
\setcounter{table}{0}
\caption{Scaling laws for the Lehnert number (defined in Eq.~\ref{equation:eq1}) obtained with different assumptions for the strength of the magnetic field force or energy densities. In the 'balance' row, $ F_L$ denotes the Lorentz force density, $F_\nu$ the viscous force density, $a_\mathrm{c}$ the Coriolis acceleration, $\ME$ and $\KE$ are the magnetic and kinetic energy densities, and $\Ro=\uc/(2\Omega\lc)$ the convective Rossby number with convective speed $\uc$ and length $\lc$. The turbulent viscosity is defined as $\nu=\uc\lc/3$. The angular frequency $\Omega$, $\uc$, and $\lc$ are outputs of the stellar code evolution STAREVOL.}
\label{table:tab1}
\end{table}

\subsection{Parametric study of the Lehnert number}
The right panel of Fig.~\ref{astoul:fig1} shows the Lehnert number squared, $\Le^2$, versus the normalized radius inside the convective zone, in the equipartition (in blue) and magnetostrophic (in black) regimes (see Table~\ref{table:tab1}). Results are shown at different evolutionary stages from the middle of the pre-main sequence ($\sim$30 Myr; in dotted lines) to the termination of the main sequence ($\sim$10.5 Gyr; in solid lines). We stress that $\Le^2$ increases with time in the equipartition regime. In this regime, there is a sharp variation of $\Le^2$ at the base and the top of the convective zone but it remains fairly flat in between, which comes about because of the radial profile of the fluid Rossby number \citep[see Fig. 4 from][]{M2016}. In the magnetostrophic regime, $\Le^2$ is rather uniform in the bulk of the convective zone, fairly constant over time from the ZAMS to the TAMS,  and decreases near the surface. Overall, we see that $\Le^2$ is always less than unity, in accordance with previous estimations \citep{LO2018,W2018}. In the following, we will quote the Lehnert number at the base of the convective zone, keeping in mind that for all prescriptions, except for magnetostrophy, $\Le$ can significantly increase towards the surface of the star. This choice corresponds to focus on the effects of large-scale magnetic fields.

\begin{figure}[ht!]
 \centering
 \includegraphics[width=0.49\textwidth,clip]{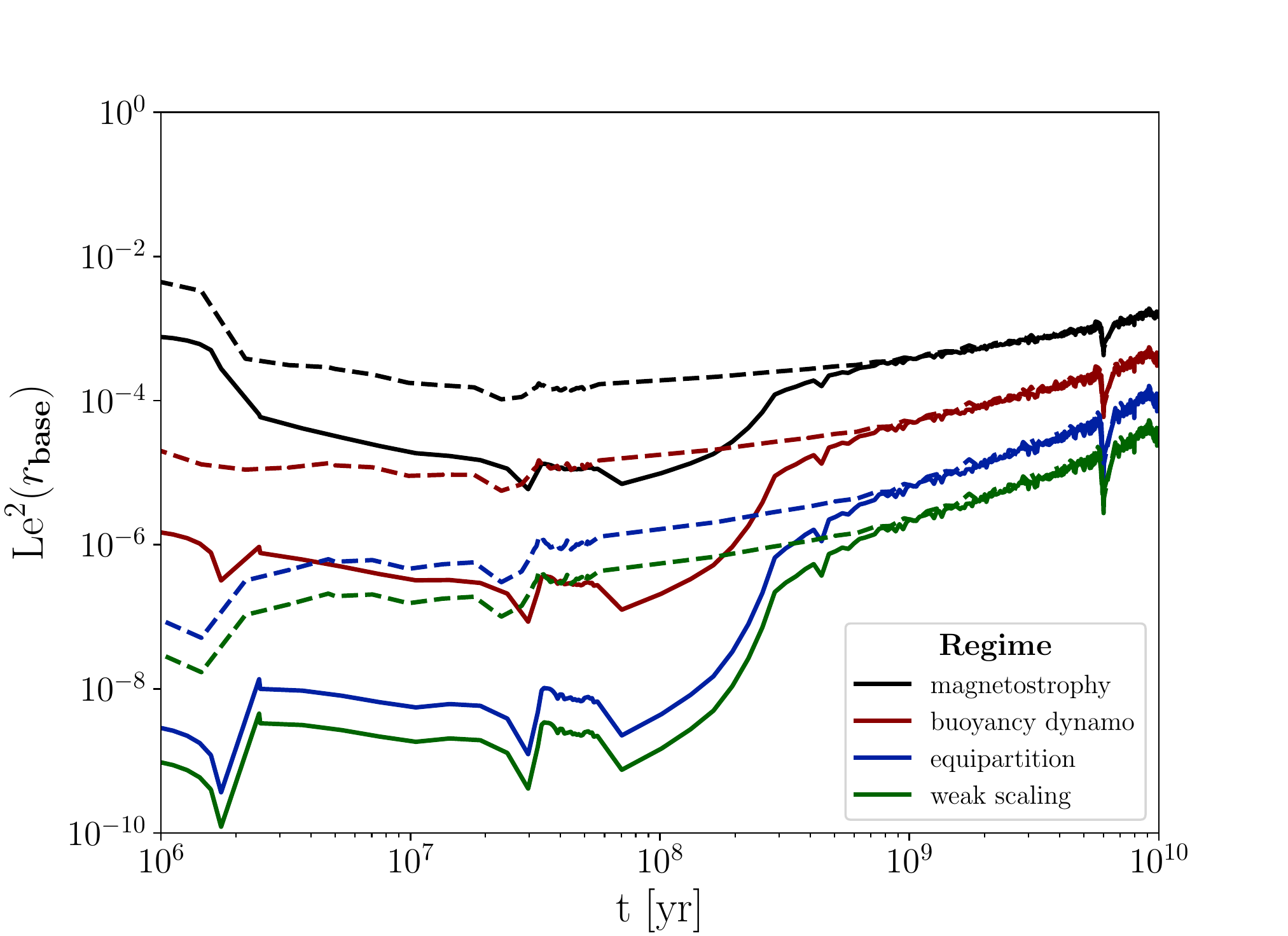}
 \includegraphics[width=0.49\textwidth,clip]{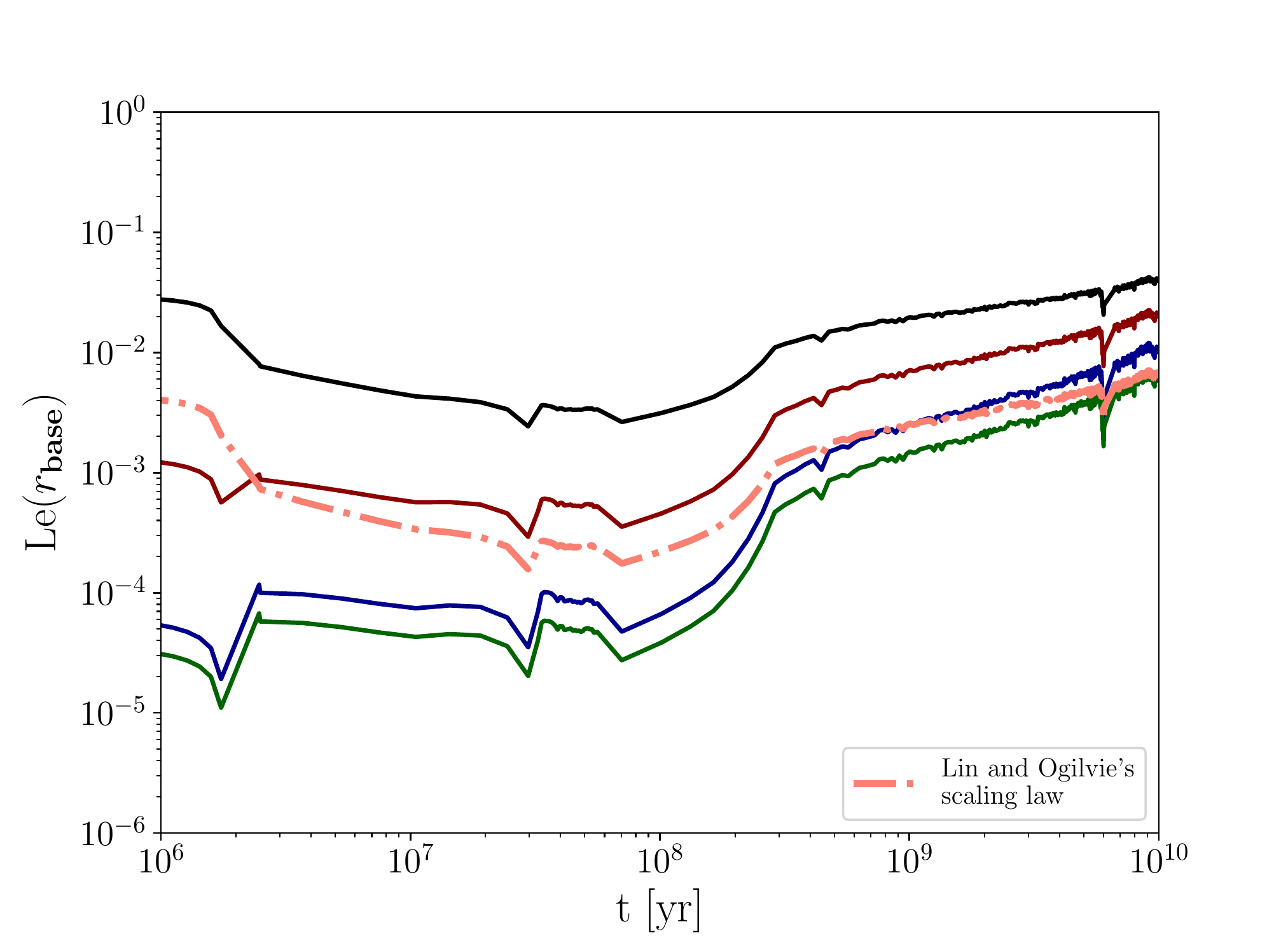}      
  \caption{{\bf Left:} Lehnert number squared at the base of the convective zone against the age of a $1M_\odot$ star, for the four different scaling laws introduced in Table~\ref{table:tab1} to estimate the star's magnetic field amplitude. Results are shown for an initial rotation period of 1.6 days (solid curves) and 9 days (dashed curves). {\bf Right:} Lehnert number at the base of the convective zone versus the age of a $1M_\odot$ star for fast initial rotation (1.6 days), and for the same scaling laws as in the left panel. The criterion derived by \citet{LO2018} to assess when Ohmic dissipation dominates over viscous turbulent diffusion of (magneto-)inertial waves,  reads $\mathrm{Le_{crit}}=\mathrm{Em}^{2/3}$, with $\mathrm{Em}=\eta/(2\Omega R_\star^2)$ the magnetic Ekman number and $\eta$ the magnetic diffusity. This scaling is overplotted by the dash-dotted curve, and we chose a turbulent magnetic diffusivity (e.g. $\eta=\uc\lc/3$).}
  \label{astoul:fig2}
\end{figure}

In the left panel of Fig.~\ref{astoul:fig2}, $\Le^2$ is displayed against time for our $1M_\odot$ model for the different scaling laws detailed in table \ref{table:tab1}. Results are shown for two different initial periods: 1.6 days (solid curves) and 9 days (dashed curves). Until about $1\ \mathrm{Gyr}$, $\Le^2$ is greater for slower initial rotation, and its variations mostly reflect those of the surface rotation period of the star as modeled in \cite{GB2013}. Beyond $1\ \mathrm{Gyr}$, $\Le^2$ increases monotonically with time as a consequence of the wind braking mechanism described by the Skumanich relationship \citep{WD1967,S1972}. One can see for instance that $\Le^2$ increases as $t^{1/2}$ in the magnetostrophic regime. 

\subsection{Does magnetic field matter for the tidal forcing in observed star-planet systems?}
\label{sec:sys}
The Doppler-shifted Rossby number $\Roh=|\sigma_\mathrm{t}|/(2\Omega)$ introduced in Eq.~(\ref{equation:eq1}) can be estimated for observed planetary systems with a solar-mass star. For the tidal frequency $\sigma_\mathrm{t}$, we assume for simplicity the case of circular and coplanar orbits, so that $\sigma_\mathrm{t}=2\Omega_\mathrm{o}-\Omega$ with $\Omega_\mathrm{o}$ the orbital frequency of the planet. The Doppler-shifted Rossby number can be recast as $\Roh=|P_{\star}/P_\mathrm{o}-1|$ with $P_{\star}$ and $P_\mathrm{o}$ the rotational and orbital periods, respectively. We see that the closer these periods are, the higher the ratio $\Le^2/\Roh$, and the larger the effect of the linearised Lorentz force when compared to those of the Coriolis acceleration on tidal forcing.
\begin{table}[h]
\centering\begin{tabular}{cccc||c}
Systems & $\mathrm{age}\ [\mathrm{Gyr}]$ & $P_\mathrm{o}\ [\mathrm{d}]$ & $P_{\star}\ [\mathrm{d}]$ & $\mathrm{Le}^2/\Roh$ \\
\hline
\rule[0.5ex]{0pt}{2ex} HAT-P-36 (b) & $6.6 \pm1.8$ & 1.3 & $15.3\pm0.4$ & $10^{-4}$ \\
WASP-5 (b) & $3.0 \pm1.4$ & 1.6 & $16.2\pm0.4$ & $9\times 10^{-5}$ \\
WASP-16 (b) & $2.3 \pm2.2$ & 3.1 & 18.5 & $10^{-4}$ \\
\end{tabular}
\caption{Estimate of $\mathrm{Le}^2/\Roh$ at the base of the convective zone for three star-planet systems with a $\sim1M_\odot$ host star and a hot Jupiter-like planet. Ages and orbital periods ($P_\mathrm{o}$) were found in \url{https://www.exoplanet.eu}. The age of the stars has been used to get the rotation period $P_{\star}$ for the star WASP-16 via STAREVOL, and $\mathrm{Le}^2$ for the magnetostrophic regime in all three systems.
The rotation periods of the host stars HAT-P-36 and WASP-5 have been found in \cite{ME2015} and \cite{MS2015}, respectively.
}\label{table:tab2}
\end{table}

In Table~\ref{table:tab2}, the ratio $\Le^2/\Roh$ is shown for actual star-planet systems of various ages and periods. These systems have stellar masses close to $1M_\odot$, and nearly circular orbits with small projected stellar obliquities. We see that $\Le^2/\Roh$ is much smaller than unity for these systems, meaning that the Lorentz force has nearly no impact on tidal forcing. This is not really surprising since these systems are far from synchronization. A state close to synchronization could be obtained for younger stars with close-in companions. It may also be obtained as a result of strong differential rotation in the star's convective envelope \citep{G2016,BB2018}.

\section{Impact of the star's magnetic field on tidal dissipation of (magneto-)inertial waves}
In Sect.~\ref{sec:sect2}, we have estimated the impact of a star's magnetic field on the effective tidal forcing that leads to wave excitation in the convective zone. However, the magnetic field can also directly affect the propagation and dissipation of these waves. For large magnetic fields, the Lorentz force can thwart the Coriolis acceleration, thereby exciting magneto-inertial waves in the convective zone \citep{W2016,LO2018}. The energy dissipation of these waves can be controlled by either Ohmic dissipation or viscous turbulent diffusion, depending on the value of the Lehnert number. More specifically, \cite{LO2018} showed that Ohmic dissipation dominates when $\Le\geq(\mathrm{Em})^{2/3}$, where $\mathrm{Em}=\eta/(2\Omega R_\star^2)$ denotes the magnetic Ekman number and $\eta$ the magnetic diffusivity. In the right panel of Fig.~\ref{astoul:fig2}, we display $\Le = (\mathrm{Em})^{2/3}$ against the star age, assuming $\eta=\nu=\uc\lc/3$ (that is, a turbulent magnetic Prandtl number of unity). Comparison with the solid curves underlines that, depending on the scaling law used to estimate the star's magnetic field, Ohmic dissipation can dominate or not. Awaiting constrained magnetic field strengths, both viscous and Ohmic dissipation should be taken into consideration.

\section{Conclusions}
We have investigated the influence of the amplitude of stellar magnetic fields on star-planet tidal interactions. We have first derived a simple analytical criterion to quantify the importance of the Lorentz force, when compared to the Coriolis acceleration, on the effective tidal forcing of (magneto-)inertial waves in the convective envelope of a solar-type star. We have used a 1D stellar evolution model to examine how the impact of the Lorentz force in the tidal forcing depends on the age of the star, its initial rotation period, and the position in the convective envelope. For coplanar and circularised star-planet systems, we have shown that the impact of large-scale magnetic fields on tidal forcing is likely small at the base of the convective envelope (that is, near the tachocline), except for near-synchronised systems. We have also used the results of our stellar evolution model to assess the importance of Ohmic diffusivity on the dissipation of (magneto-)inertial waves, by using the criterion derived by \citet{LO2018}. Our results indicate that Ohmic and viscous turbulent dissipations have similar magnitude at the base of the convective zone in solar-mass stars, so that both diffusion processes should be taken into account for the calculation of the dynamical tide. In a paper in preparation, we will expand and detail these preliminary results, and will also consider lower-mass stars.
Moreover, the study of the impact of magnetic field on tides must be combined with the analysis of planet-star interactions through magnetic couples \citep{SB2017}.

\begin{acknowledgements}
A. Astoul, S. Mathis and E. Bolmont acknowledge funding by the European Research Council through the ERC grant SPIRE 647383. The authors acknowledge the PLATO CNES funding at CEA/IRFU/DAp and IRAP, and the CNRS/INSU PNP funding.
\end{acknowledgements}

\bibliographystyle{aa}
\bibliography{astoul.bib}
\end{document}